\documentstyle[epsfig,amsmath,amssymb,graphicx,tabularx]{elsart}
\newcounter{saveeqn}

\def\gsimeq{\,\,\raise0.14em\hbox{$>$}\kern-0.76em\lower0.28em\hbox  
{$\sim$}\,\,}  
\def\lsimeq{\,\,\raise0.14em\hbox{$<$}\kern-0.76em\lower0.28em\hbox  
{$\sim$}\,\,}  
\def\beqy{\begin{eqnarray}}
\def\eeqy{\end{eqnarray}}
\def\bmlet{\begin{mathletters}}
\def\emlet{\end{mathletters}}
\newcommand{\rhob}{\mbox{\boldmath$\rho'$}}
\newcommand{\hob}{\mbox{\boldmath$\cal{H}'$}}
\begin{document}
 \begin{frontmatter}  
 
\title{Microscopic HFB+QRPA predictions of dipole strength for astrophysics
applications}
\author{S. Goriely$^1$, E. Khan$^2$, M. Samyn$^1$}
\address{
$^1$Institut d'Astronomie et d'Astrophysique, ULB - CP226, 1050 Brussels, Belgium \\
$^2$ Institut de Physique Nucl\'eaire, IN$_{2}$P$_{3}$-CNRS, 91406 Orsay,
France}
%
\begin{abstract}
Large-scale QRPA calculations of the E1 strength are performed on top of HFB
calculations in order to derive the radiative neutron capture cross sections
for the whole nuclear chart. The spreading width of the GDR is taken into
account by analogy with the second-RPA (SRPA) method. The accuracy of
HFB+QRPA model based on various Skyrme forces with different pairing
prescription and parameterization is analyzed. It is shown that the present
model allows to constrain the effective nucleon-nucleon interaction with the
GDR data and to provide quantitative predictions of dipole strengths.

\end{abstract}
\end{frontmatter}

\section{Introduction}
About half of the nuclei with $A>60$ observed in nature are formed by
the so-called rapid neutron-capture process (or r-process) of nucleosynthesis, occurring
in explosive stellar events.  
 The r-process is believed to take place in environments characterized by high neutron
densities ($N_n \gsimeq 10^{20}~{\rm cm^{-3}}$), so that successive
neutron captures proceed into neutron-rich regions well off the
$\beta$-stability valley forming exotic nuclei that cannot be produced and therefore
studied in the laboratory.  If the temperatures or the neutron densities characterizing
the r-process are low enough to break the $(n,\gamma)-(\gamma,n)$ equilibrium, the
r-abundance distribution depends directly on the neutron capture rates by the
so-produced exotic neutron-rich nuclei \cite{go98}. The neutron capture rates
are commonly evaluated within the framework  of the statistical model of
Hauser-Feshbach (although the direct capture contribution can play an important role for
such exotic nuclei).
This model makes the fundamental assumption that the capture process takes place with the
intermediary formation of a compound nucleus in thermodynamic equilibrium. In this
approach, the Maxwellian-averaged $(n,\gamma)$ rate at temperatures of relevance in
r-process environments strongly depends on the electromagnetic interaction, i.e the
photon de-excitation probability. The well known challenge of
understanding the r-process abundances thus requires that one be able to make
reliable extrapolations of the E1-strength function out towards the neutron-drip line.
To put the description of the r-process on safer grounds, a great effort
must therefore be made to improve the reliability of the nuclear
model. Generally speaking, the more microscopic the underlying theory, the greater will
be one's confidence in the extrapolations out towards the neutron-drip line, provided,
of course, the available experimental data are also well fitted.

Large scale prediction of E1-strength functions are usually performed using
phenomenological Lorentzian models of the giant dipole resonance (GDR) \cite{go98}.
Several refinements can be made, such as the energy dependence of the width and its
temperature dependence
\cite{go98,mc81,ka83,ko87} to describe all available experimental data. The Lorentzian
GDR approach suffers, however, from shortcomings of various
sorts. On the one hand, it is unable to predict the enhancement of the E1 strength at
energies around the neutron separation energy demonstrated by various experiments, such
as the nuclear resonance fluorescence. On the other hand, even if a Lorentzian-like
function provides a suitable representation of the E1 strength for stable nuclei, the
location of its maximum and its width remain to be predicted from some systematics or
underlying model for each nucleus. For astrophysical applications, these properties have
often been obtained from a droplet-type model \cite{my77}. This approach clearly lacks
reliability when dealing with exotic nuclei, as already demonstrated by
\cite{ca97,go02}. Recently an attempt was made to derive microscopically the E1 strength
for the whole nuclear chart \cite{go02}. The dipole response was calculated with the
Quasiparticle Random Phase Approximation (QRPA) on top of Hartree-Fock+BCS (HFBCS)
description \cite{kh00}. The only input of this approach was the Skyrme effective
interaction injected in the HFBCS model. These microscopic calculations predicted the
presence of a systematic low-lying component in the E1 strength for very neutron-rich
nuclei. This low-lying component influences the neutron capture rate, especially if
located in the vicinity of the neutron separation energy $S_n$.  

In our previous HFBCS and QRPA microscopic approach \cite{go02}, the pairing correlation
in the BCS model was determined assuming a simple constant-gap pairing interaction. In
addition,  in the case of the highly neutron-rich nuclei that are of particular interest
in the context of the r-process, the validity of the BCS approach to pairing is
questionable, essentially because of the role played by the continuum of
single-particle neutron states (see \cite{do96},  and references therein).
Therefore the impact of the newly-derived E1-strength functions on the cross section
prediction could only be evaluated qualitatively. It was found that the radiative
neutron capture cross sections by neutron-rich nuclei were systematically increased by
the HFBCS+QRPA calculations \cite{go02} with respect to the one obtained using
Lorentzian-like strength functions. Predictions with different forces have been
compared, but no conclusions could be drawn regarding their intrinsic quality to predict
the E1 strength. The final large-scale HFBCS+QRPA calculations performed in \cite{go02}
were obtained on the basis of the Skyrme force denoted SLy4 \cite{ch98}.

In the present paper we calculate the dipole strength with one of the most accurate and
reliable microscopic model available to date, namely the Hartree-Fock-Bogoliubov (HFB)
and QRPA models
\cite{gr01,kh02}. As recalled in Sect.~2.1, the ground state is described
within the HFB model. Effective interactions of the Skyrme type
characterized by different values of the nucleon effective mass and
prescriptions for the pairing interaction are considered. The collective
GDR mode is obtained by QRPA calculations on top
of the HFB calculations, as described in Sect.~2.2. The residual interaction
is derived self-consistently from the nucleon-nucleon effective interaction,
which is the only input of the HFB calculation. To describe the damping of
the collective motions on microscopic grounds, the second-RPA (SRPA)
described by \cite{dr90} is adopted (Sect.~2.3). This approach strongly
improves the reliability of the predictions by eliminating the
phenomenological spreading of the QRPA strength determined in our previous
HFBCS+QRPA calculations \cite{go02}. This new approach allows us to
determine on a more quantitative and reliable ground the photoabsorption
cross section and consequently to judge the ability of the forces to reproduce
experimental data. 
In order to select the most adequate interaction for the E1-strength calculation, the
HFB+QRPA prediction are compared with photoabsorption data for 48 spherical nuclei
\cite{di89,iaea00} (Sect.~3).  The HFB+QRPA predictions are further compared, in Sect.~3,
with low-energy experimental data and generalized to include deformation effects in a
phenomenological way. All these drastic improvements compared to the previous HFBCS and
QRPA models allow to provide quantitative predictions of the E1-strength function, also
for exotic neutron-rich nuclei (Sect.~4). The predicted GDR strengths are used to
estimate all the radiative neutron capture rates of relevance for nucleosynthesis
applications (Sect.~4).

\section{HFB+QRPA calculation of the E1-strength function}

The long-term goal of microscopic models is to describe on the same ground a
wide variety of nuclear structure properties (in particular, magicity and pairing
correlations in open-shell nuclei) for both stable and exotic nuclei. The HFB and QRPA
models allows to treat, in a self-consistent way, pairing effects on the ground state
as well as collective excitations for nuclei ranging from the valley of stability to the
drip-line. The QRPA considers nuclear excitation as a collective superposition of two
quasiparticle (qp) states built on top of the HFB ground state \cite{rs80}. This
collective aspect of the excitation makes the QRPA an accurate tool to investigate the
E1-strength function, in both closed and open shell nuclei.

\subsection{HFB Calculations}

The HFB calculations considered in the present work are fully detailed in
\cite{ms02,sg02,ms03,ms03b}. They are based on the conventional Skyrme force of the form
\begin{eqnarray}
\label{eq_sky}
v_{ij} & = & t_0(1+x_0P_\sigma)\delta({{\bf r}_{ij}})
+t_1(1+x_1P_\sigma)\frac{1}{2\hbar^2}\{p_{ij}^2\delta({{\bf r}_{ij}})
+h.c.\}\nonumber\\
& &+t_2(1+x_2P_\sigma)\frac{1}{\hbar^2}{\bf p}_{ij}.\delta({\bf r}_{ij})
 {\bf p}_{ij}
+\frac{1}{6}t_3(1+x_3P_\sigma)\rho^\gamma\delta({\bf r}_{ij})\nonumber\\
& &+\frac{i}{\hbar^2}W_0(\mbox{\boldmath$\sigma_i+\sigma_j$})
{\bf .p}_{ij}\times\delta({\bf r}_{ij}){\bf p}_{ij}  \quad .
\end{eqnarray} 
The pairing force acting between 
like nucleons is treated in the full Bogoliubov framework with a
$\delta$-function pairing  force of the form \cite{ms03,ms03b,gar99}
\begin{equation}
v_{\pi}(\mbox{\boldmath$r$}_{ij})=
V_{\pi q}~\left[1-\eta
\left(\frac{\rho}{\rho_0}\right)^\alpha\right]~
\delta(\mbox{\boldmath$r$}_{ij}) \quad ,
\label{eq_surf}
\end{equation}
where $\rho$ is the density and $\rho_0$ its equilibrium value in symmetric nuclear
matter. Two types of pairing forces are considered here, a
volume  density-independent  force characterized by $\eta=0$ and a volume plus surface
(i.e density-dependent) force with $\eta=0.45$ and
$\alpha=0.47$. This latter prescription originates from the calculations of the
pairing gap in infinite nuclear matter at different densities performed by Garrido et
al. \cite{gar99} using a ``bare"  or ``realistic" nucleon-nucleon interaction. This
density-dependent pairing has also been found to be compatible with experimental
nuclear masses by \cite{ms03}, provided the space of single-particle states over which
such a pairing force is allowed to act is truncated to about
$\varepsilon_{\Lambda}\simeq 15$~MeV around the Fermi energy. Note finally that in the
present approach the strength parameter
$V_{\pi q}$ is allowed to be different for  neutrons and 
protons, and also to be slightly stronger for an odd number of nucleons 
($V_{\pi q}^-$) than for an even number ($V_{\pi q}^+$), i.e., 
the pairing force between neutrons, for example, depends on whether $N$ is 
even or odd. For odd-$A$ and odd-odd nuclei, the blocking approximation is used, as
detailed in \cite{ms02}.

Based on this Skyrme-HFB approach, a number of effective forces have been determined
recently \cite{sg02,ms03,ms03b}, the  parameters of the underlying forces being fitted
{\it exclusively} to all  the 2135 available experimental masses \cite{aw01}, with some
additional constraints regarding the stability of neutron matter and the
incompressibility of nuclear matter. The parameters corresponding to these six forces
named BSk2-BSk7 are summarized in Table~\ref{tab_sky}. Are also included in
Table~\ref{tab_sky} the effective isoscalar ($M^*_s$), isovector ($M^*_v$) nucleon mass
and the root-mean-square (rms) deviations
$\sigma$ between the measured and estimated masses for the 2135 nuclei with
$Z,N \geq 8$. More details about these forces can be found in
\cite{sg02,ms03,ms03b}. The major differences between these six forces are found in the
density-dependence of the pairing force and the adopted isoscalar effective nucleon
mass. While BSk2 and BSk3 have been built without constraining the effective mass
(leading to a value of $M^*_s/M \gsimeq 1.04$),
 BSk4 and BSk5 are constrained by $M^*_s/M = 0.92$, as inferred from the extended
Br\"uckner-Hartree-Fock calculations of asymmetric nuclear matter \cite{zbl99} and BSk6
and BSk7 by $M^*_s/M=0.8$, as inspired from the more traditional symmetric
nuclear-matter calculations (e.g \cite{jlm76}). The mass-data fits with the BSk4--7
interaction are almost as good as those obtained with BSk2--3, in which
$M_s^*/M$ is unconstrained, so that such a mass fit cannot be used 
to discriminate between the different Skyrme forces and the
corresponding optimal choice for the nucleon effective mass or the pairing interaction.
For comparison purposes, we also consider the SLy4 Skyrme force
\cite{ch98} used in our previous HFBCS+QRPA calculation
\cite{go02}. The SLy4 parameters are given in Table~\ref{tab_sky}, the pairing
interaction corresponding to the one determined in \cite{kh02} (the mass rms deviation is
however not available for the SLy4 force).

 \begin{table}[h]
 \centering
 \caption{Some properties of the Skyrme forces BSk2-BSk7 and
SLy4 (see text for more details). The last line corresponds to the rms deviation
between predicted and experimental masses for the full set of 2135 spherical and
deformed nuclei.}
\label{tab_sky}
\vspace{.5cm}
 \begin{tabular}{|c|ccccccc|}
 \hline
  & BSk2 & BSk3 & BSk4 & BSk5 & BSk6 & BSk7 & SLy4 \\
 \hline
  $t_0$ {\scriptsize [MeV fm$^3$]} & -1790.63 & -1755.13      & -1776.94 &
-1778.89 & -2043.32 & -2044.25 & -2488.91\\
  $t_1$ {\scriptsize [MeV fm$^5$]} & 260.996& 233.262      & 306.884 & 312.727 & 382.127 
& 385.973 & 486.818 \\
  $t_2$ {\scriptsize [MeV fm$^5$]} & -147.167& -135.284 & -105.670 & -102.883 &
-173.879 & -131.525 & -546.395 \\
  $t_3$ {\scriptsize [MeV fm$^{3+3\gamma}$]} & 13215.1&13543.2 & 12302.1 & 12318.37 &
12511.7 & 12518.8 & 13777.0 \\
  $x_0$             & 0.4990 &  0.4766      &  0.5426 & 0.4445 & 0.7359 &
0.7292 & 0.834\\
  $x_1$             & -0.0898   & -0.0326      & -0.5352 & -0.4887 & -0.7992 & 
-0.9323 & -0.344\\
  $x_2$  & 0.2244  &  0.4704     &  0.4947 & 0.5846 &-0.3590 & -0.0501 &
-1.0000 \\
  $x_3$             & 0.5157  &  0.4225    &   0.7590 & 0.5693 & 1.2348 &
1.2363 & 1.3540\\
  $W_0$ {\scriptsize [MeV fm$^5$]} & 119.05 & 116.07  & 129.50 & 130.70 & 142.38 & 146.93
& 123.00 \\
  $\gamma$          & 0.3433   & 0.3612 & 1/3 & 1/3 & 1/4 & 1/4 & 1/6 \\
  $V^+_{\pi n}$ {\scriptsize [MeV fm$^3$]} & -238 & -359  & -273 & -429 & -321 & -505 &
-395\\
  $V^-_{\pi n}$ {\scriptsize [MeV fm$^3$]} & -265 &-407  & -289 & -463 & -325 & -514 &
-395\\
  $V^+_{\pi p}$ {\scriptsize [MeV fm$^3$]} & -247 &-365  & -285 & -447 & -338 & -531&
-395\\
  $V^-_{\pi p}$ {\scriptsize [MeV fm$^3$]} & -278 &-413  & -302 & -483 & -341 & -541&
-395\\
  $\eta$                      & 0 & 0.45 & 0 & 0.45 & 0 & 0.45 & 1.00\\
  $\alpha$                    & 0  & 0.47 & 0  & 0.47 & 0 & 0.47 & 1.50 \\
  $\varepsilon_{\Lambda}${\scriptsize [MeV]} & 15 & 14 & 16  & 16 & 17 & 17 & 60\\

  $M^*_s/M$ & 1.04 & 1.12 & 0.92 & 0.92 & 0.80 & 0.80 & 0.69\\
  $M^*_v/M$ &0.86 & 0.89 & 0.85 & 0.84 & 0.86 & 0.87 & 0.80\\
 \hline
 $\sigma${\scriptsize [MeV]} & 0.674 & 0.656 & 0.680   & 0.675  &
0.686 & 0.676 & -- \\
 \hline
 \end{tabular}
 \end{table}

\subsection{QRPA Calculations}

 The E1-strength QRPA calculations are performed on top of the HFB results.
 The derivation of the QRPA response is detailed in \cite{kh02}, using
 Green's function formalism. The QRPA response is obtained from the
 time-dependent HFB equations \cite{rs80}:
\begin{equation}\label{eq:tdhfb}
i\hbar\frac{\partial{\cal R}}{\partial t}=[{\cal H}(t) +
{\cal F}(t),{\cal R}(t)]
\end{equation}
where ${\cal R}$, ${\cal H}$ are the time-dependent generalized density and
HFB Hamiltonian respectively, and ${\cal F}$ the external oscillating field.
In the small amplitude limit the time-dependent HFB equations become:
\begin{equation}\label{eq:lin}
        \hbar\omega{\cal R}'=[{\cal H}',{\cal R}^0] + [{\cal H}^0,{\cal
	        R}']+[F,{\cal R}^0]
\end{equation}
where $'$ stands for the perturbed quantity. 
The variation of the generalized density ${\cal R}'$ is expressed in
term of 3 quantities, namely   $\rho'$, $\kappa'$ and $\bar{\kappa}'$,
which are written as a column vector  
\begin{equation}\label{eq:rhodef}
  \rhob        =\left(
	\begin{array}{c}
	\rho' \\
         \kappa' \\
	 \bar{\kappa}' \\
        \end{array}
	\right) \quad .
	\end{equation}
Thus, at variance with the RPA, where one needs to know only the change of
the particle-hole (ph) density ($\rho'$), in QRPA one should calculate the variation of
three basic quantities (Eq. \ref{eq:rhodef}). It should be noted that in
the three dimensional space introduced above, the first dimension represents
the ph subspace, the second the particle-particle (pp) one,
and the third the hole-hole (hh) one. The response matrix has 9 coupled
elements in QRPA, compared to one in the RPA formalism.

The variation of the HFB Hamiltonian is expressed in terms of the
second derivatives of the HFB energy functional ${\cal
E}$[$\rho,\kappa,\bar{\kappa}$] with respect to the densities 
\begin{equation}\label{eq:hvar}
\hob=\bf{V}\rhob ,
\end{equation}
where $\bf{V}$ is the residual interaction matrix, namely :
\begin{equation}\label{eq:vres}
{\bf{V}}^{\alpha\beta}({\bf r}\sigma,{\bf r}'{\sigma}')=
\frac{\partial^2{\cal E}}{\partial{\bf{\rho}}_\beta({\bf r}'{\sigma}')
\partial{\bf{\rho}}_{\bar{\alpha}}({\bf r}\sigma)},~~~\alpha,\beta = 1,2,3.
\end{equation}
Here, the notation $\bar{\alpha}$ means that whenever $\alpha$ is 2 or 3
then $\bar{\alpha}$ is 3 or 2.

The quantity of interest is the QRPA Green's function $\bf{G}$, which relates
the perturbing external field to the density change by
\begin{equation}\label{eq:g}
\rhob=\bf{G}\bf{F}~.
\end{equation}
Replacing the above three equations in Eq.~(\ref{eq:lin}), yields the
so-called Bethe-Salpeter equation 
\begin{equation}\label{eq:bs}
\bf{G}=\left(1-\bf{G}_0\bf{V}\right)^{-1}\bf{G}_0=\bf{G}_0+\bf{G}_0\bf{V}\bf{G}
\end{equation}
corresponding to a set of 3x3 coupled equations.
In Eq.~(\ref{eq:bs}), the unperturbed Green's function $\bf{G}_0$ is defined by :
\begin{equation}\label{eq:g0}
{\bf{G}_0}^{\alpha\beta}({\bf r}\sigma,{\bf r}'{\sigma}';\omega)=
\sum_{ij} \frac{{\cal U}^{\alpha 1}_{ij}({\bf r}\sigma)
\bar{{\cal U}}^{*\beta 1}_{ij}({\bf r}'\sigma')}{\hbar\omega-(E_i+E_j)+i\eta}
-\frac{{\cal U}^{\alpha 2}_{ij}({\bf r}\sigma)
\bar{{\cal U}}^{*\beta 2}_{ij}({\bf r}'\sigma')}{\hbar\omega+(E_i+E_j)+i\eta} \quad ,
\end{equation}
where $E_i$ are the single qp energies and ${\cal U}_{ij}$ are 3 by 2
matrices calculated from the $U$ and $V$ HFB wave functions \cite{kh02}. 

In the case of transitions from the ground state to excited states within
the same nucleus, only the (ph,ph) component of $\bf{G}$ plays a role. If the
interaction does not depend on spin variables the strength function is thus
given by 
\begin{equation}\label{eq:stren}
S(\omega)=-\frac{1}{\pi}Im \int F^{11*}({\bf r}){\bf{
G}}^{11}({\bf r},{\bf r}';\omega)F^{11}({\bf r}') 
d{\bf r}~d{\bf r}' \quad .
\end{equation}

The QRPA calculations are performed assuming the spherical symmetry. The residual
interaction is derived from the interaction used in the HFB calculation (Eq.
\ref{eq:vres}). The residual interaction corresponding to the
velocity-dependent terms of the Skyrme force is approximated in the (ph,ph)
subspace by its Landau-Migdal limit \cite{ba75}. All the qp states are
included up to an energy cutoff of 60 MeV, allowing pairs of
qp energy up to 120 MeV. The strength distribution is calculated up to a maximum
transition energy $\omega_{max}$=30 MeV  with a step of 100 keV and an averaging
width $\eta$=150 keV. In a fully consistent calculation the spurious
center-of-mass state should come out at zero energy. Because of the
Landau-Migdal form of the interaction adopted here, the consistency between
mean field and residual qp interaction is broken and the isoscalar $J^{\pi}
= 1^{-}$ spurious state becomes imaginary. We cure this defect by
renormalizing the residual interaction by a factor $\alpha$ on the nuclei of
interest. The spurious state comes out at zero energy with typical $\alpha$
values between 0.85 and 1. 

\subsection{Second-RPA damping}

The QRPA provides a reliable description of the GDR centroid
and the fraction of the energy-weighted sum rule (EWSR) exhausted by the E1 mode.
However, it is necessary to go beyond this approximation scheme in order to describe the
damping properties of the collective motion. The GDR is known experimentally to have a
large energy width and therefore a finite lifetime. Different microscopic theories exist
(see e.g \cite{dr90,co01,sc01}).

In the previous HFBCS and QRPA calculation, the QRPA strength was
folded by an arbitrary Lorentzian to generate the experimentally observed
GDR width. We propose to describe the width on more microscopic grounds, by
calculating it in the second-RPA (SRPA) framework
\cite{dr90}. The SRPA allows to take into account the spreading width due to
the 2p-2h excitations. Formally, self-energy insertions on particle and hole
lines spread the resonances and shift their centroids. In practice the ph
QRPA strength is folded by a Lorentzian function representing the self-energy
operator \cite{dr90,sm88}
\begin{equation}
f_L(E',E)= \frac{1}{2\pi} ~\frac{\Gamma(E)}
{(E'-E-\Delta(E))^2+\Gamma(E)^2/4} \quad .
\label{loren}
\end{equation}
where $E'$ is the excitation energy of the ph-QRPA response, $\Delta(E)$
and $\Gamma(E)$ the real and complex part of the self-energy, respectively
\cite{dr90}. The energy dependent width $\Gamma(E)$ can be calculated from the
measured decay width of particle ($\gamma_p$) and hole ($\gamma_h$) states
\cite{dr90}
\begin{equation}
\Gamma(E)=\frac{1}{E} \int_0^E{d\epsilon \left [
\gamma_p(\epsilon)+\gamma_h(\epsilon-E) \right ](1+C_{ST})} \quad .
\label{width}
\end{equation}
The real part $\Delta(E)$ of the self-energy is 
obtained from $\Gamma(E)$ by a dispersion relation \cite{dr90}. This empirical way of
determining $\Gamma(E)$ has the advantage of including, in principle, contributions from
the excitation beyond 2p-2h
\cite{sm88}. The resulting resonance width can therefore be compared with
experimental data, such as photoabsorption cross sections. The interference factor
$C_{ST}$ in Eq. (\ref{width}) is due to the screening corrections of the exchange
interaction which can interfere destructively with self-energy diagrams
\cite{dr90}. The microscopic evaluation of this factor is delicate. In practice,
$C_{ST}$ can be adjusted phenomenologically to reproduce experimental data
\cite{wa03}, the same value being used for all nuclei. In our approach, it will be tuned
to reproduce at best the 48 experimental GDR widths and peak energies in spherical nuclei
\cite{di89,iaea00}, as shown in the next section. 

\section{Comparison with experimental data}

\subsection{Photoabsorption data}
\begin{figure}
\centerline{\epsfig{figure=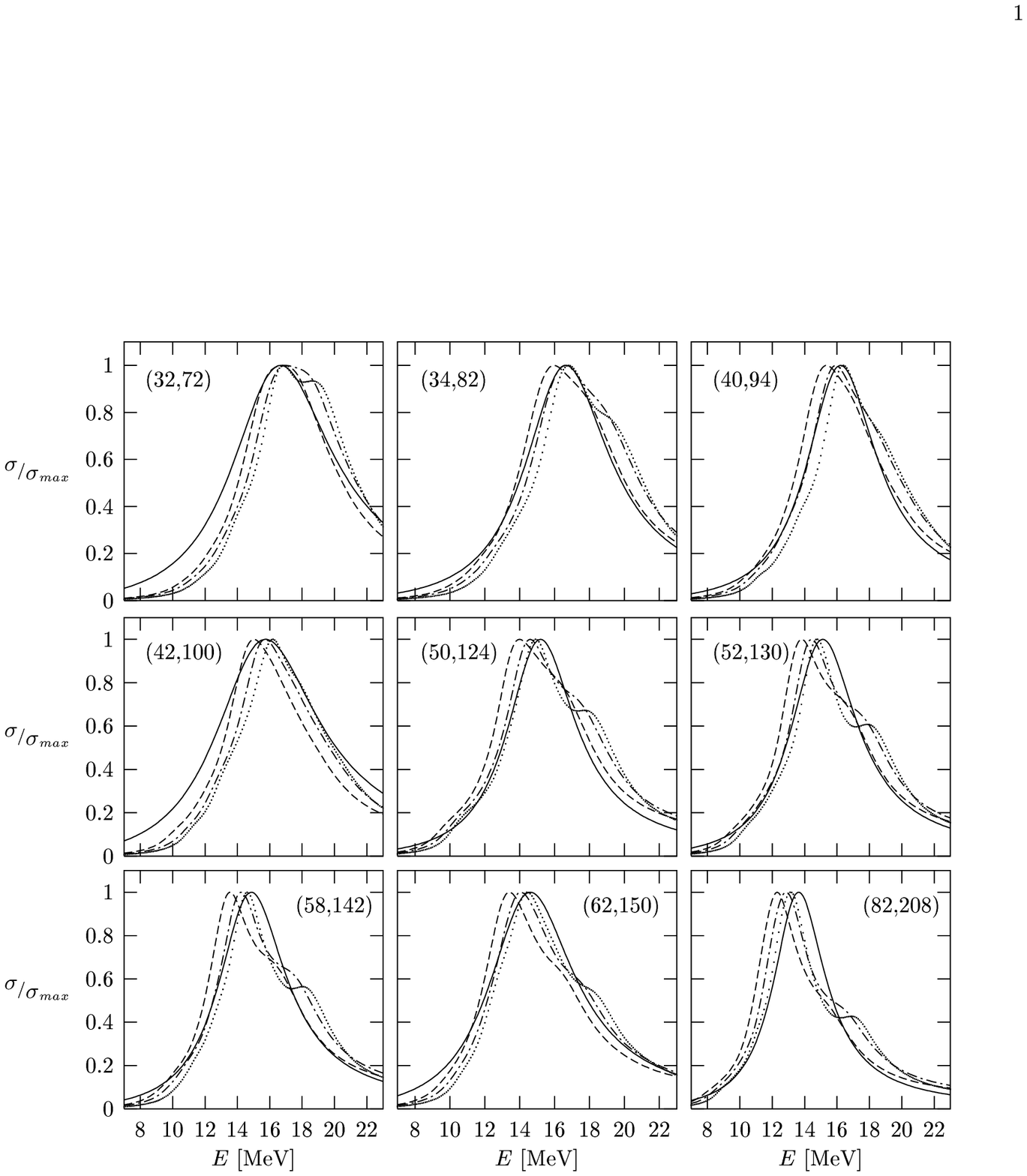,height=14.0cm,width=14cm}}
\caption{Comparison between the experimental photoabsorption cross
section approximated by a simple Lorentzian curve (solid line) in the vicinity of the
peak energy 
\cite{di89} and the QRPA calculations obtained with the Skyrme
forces BSk2 (dashed line), BSk4 (dash-dot line) and BSk6 (dotted line) for 9
representative spherical nuclei given by $(Z,A)$. All cross sections are normalized to
a peak cross section of unity.}
\label{fig_gdr1}
\end{figure}

\begin{figure}
\centerline{\epsfig{figure=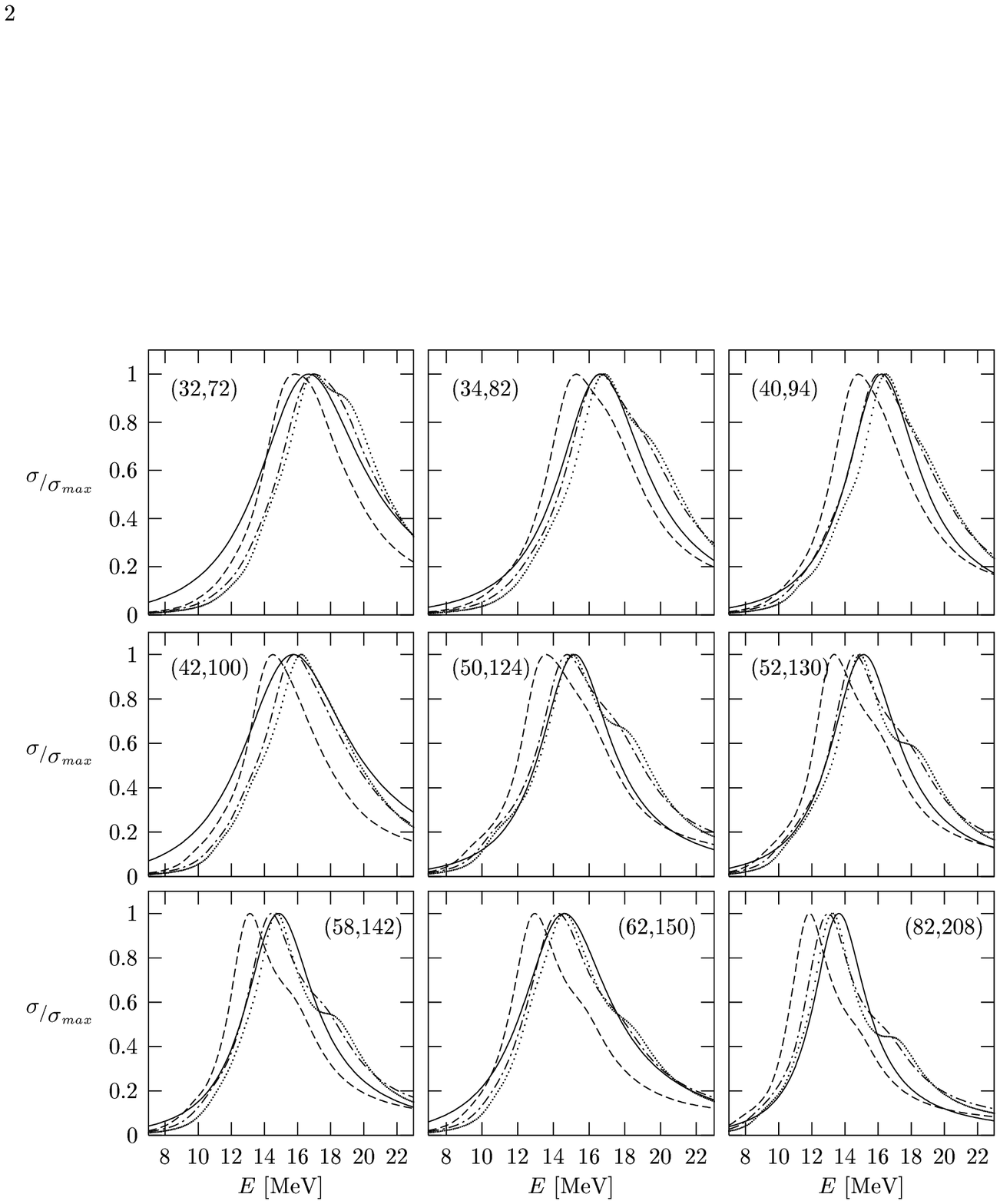,height=14.0cm,width=14cm}}
\caption{Same as Fig.~\ref{fig_gdr1} with the Skyrme
forces BSk3 (dashed line), BSk5 (dash-dot line) and BSk7 (dotted line).}
\label{fig_gdr2}
\end{figure}

\begin{figure}
\centerline{\epsfig{figure=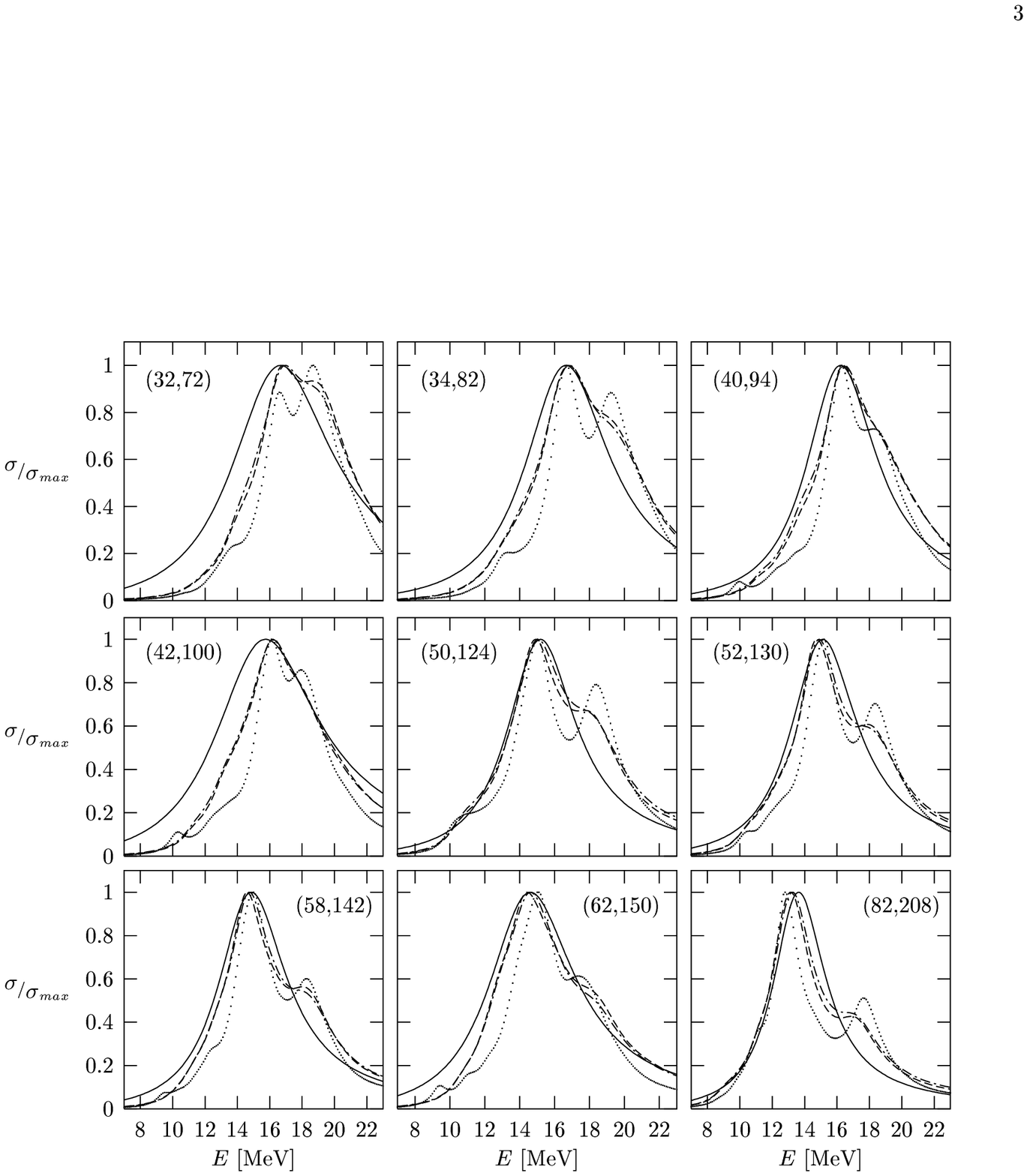,height=14.0cm,width=14cm}}
\caption{Same as Fig.~\ref{fig_gdr1} with the Skyrme
forces BSk6 (dashed line), BSk7 (dash-dot line) and SLy4 (dotted line).}
\label{fig_gdr3}
\end{figure}

Photo-induced reaction cross sections have been compiled by
\cite{di89,iaea00} and represent the most relevant and reliable source of experimental
data with which HFB+QRPA predictions can be compared. These compilations provide the GDR
parameters, i.e the peak energy, peak cross
section and the full width at half maximum, observed in photonuclear reactions measured
by bremsstrahlung, quasimonoenergetic and tagged photons for about 84 nuclei. Only
photoabsorption data for the 48 spherical nuclei are considered at this stage, in order
to free us from deformation effects. Figs. \ref{fig_gdr1}-\ref{fig_gdr3} compare
the normalized experimental photoabsorption cross section with the QRPA predictions
obtained with the 7 different Skyrme forces of Table~\ref{tab_sky}. For each force, the
interference factor C$_{ST}$ is determined to reproduce at best the position of the peak
energy and the full width at half maximum, simultaneously. As already stressed, the
interference factor influences not only the GDR width, but also 
shifts the energy peak to higher energies.  At energies around $E \simeq 15$~MeV, the
SRPA shifts the peak energy by approximately $\Delta(E)\simeq 5\times(1+C_{ST})$~MeV
upwards.  For values $C_{ST} \gsimeq -0.5$, the
GDR broadening becomes too large and incompatible with experimental photoabsorption data,
while for values
$C_{ST} \lsimeq -0.8$, the fine structure inherent to the 1p-1h QRPA estimate is not
smeared out enough by the SRPA and again incompatible with experimental 
data. The final $C_{ST}$ values are given in Table~\ref{tab_cst}, as well as the rms
deviations regarding the peak energy $E_{GDR}$ and the full width at half maximum
$\Gamma_{GDR}$.

 \begin{table}[h]
 \centering
 \caption{Interference factors adopted in the SRPA for the Skyrme forces BSk2-BSk7
and SLy4. Also given are the rms deviations for the peak energy $E_{GDR}$ and the full
width at half maximum $\Gamma_{GDR}$ relative to the experimental values for the 48
spherical nuclei compiled in \cite{di89,iaea00}.}
\label{tab_cst}
\vspace{.5cm}
 \begin{tabular}{|c|ccccccc|}
 \hline
  & BSk2 & BSk3 & BSk4 & BSk5 & BSk6 & BSk7 & SLy4 \\
 \hline
  $C_{ST}$ & -0.55 & -0.55 & -0.57 &
-0.55 & -0.66 & -0.63 & -0.76\\
  $\sigma(E_{GDR})$ {\scriptsize [MeV]} & 1.217 & 1.664 & 0.702 &
0.630 & 0.541 & 0.445 & 0.847\\
  $\sigma(\Gamma_{GDR})$ {\scriptsize [MeV]} & 1.115 & 0.924 & 1.269 &
1.135 & 1.178 & 1.169 & 1.396\\
 \hline
 \end{tabular}
 \end{table}

As illustrated in Figs. \ref{fig_gdr1}-\ref{fig_gdr3} and Table~\ref{tab_cst}, the
prediction of the GDR parameters is force-dependent.
As far as the position of the peak energy is concerned, most of the forces overpredict
the peak energy of light nuclei and underpredict it for the heavier species. The best
agreement is found for the BSk6-7 forces which still overestimate $E_{GDR}$ for the
lightest elements, but give an excellent agreement for $Z\gsimeq 40$ elements. It can be
concluded that, within the present HFB+QRPA model, Skyrme forces need to have a low
effective nucleon mass $M^*_s/M \simeq 0.8$ to correctly predict the GDR
characteristics. An effective mass as low as the one used in SLy4 requires a particularly
low value of the interference factor (see Table~\ref{tab_cst}) which simultaneously give
rise to fine structure effects not observed experimentally (see Fig.~\ref{fig_gdr3}).

 Interestingly, the density dependence of the pairing interaction
has a minor impact on the prediction of the E1-strength function. Almost no
differences are found among the 2 couples of forces BSk4 vs BSk5 or BSk6 vs BSk7 (see in
particular Fig.~\ref{fig_gdr3}). Finally, note that the SRPA effect to shift the energy
peak was not taken into account by the phenomenological Lorentzian damping adopted in
our previous HFBCS+QRPA work \cite{go02}, and obviously modifies the conclusion drawn
regarding the ability of the SLy4 force to predict the location of the GDR. The
agreement found here for the SLy4 interaction (Fig.~\ref{fig_gdr3}) is worse than it
used to be in
\cite{go02} where an rms deviation of 0.457~MeV was obtained on the centroid energy for
the same sample of spherical nuclei.


\begin{figure}
\centerline{\epsfig{figure=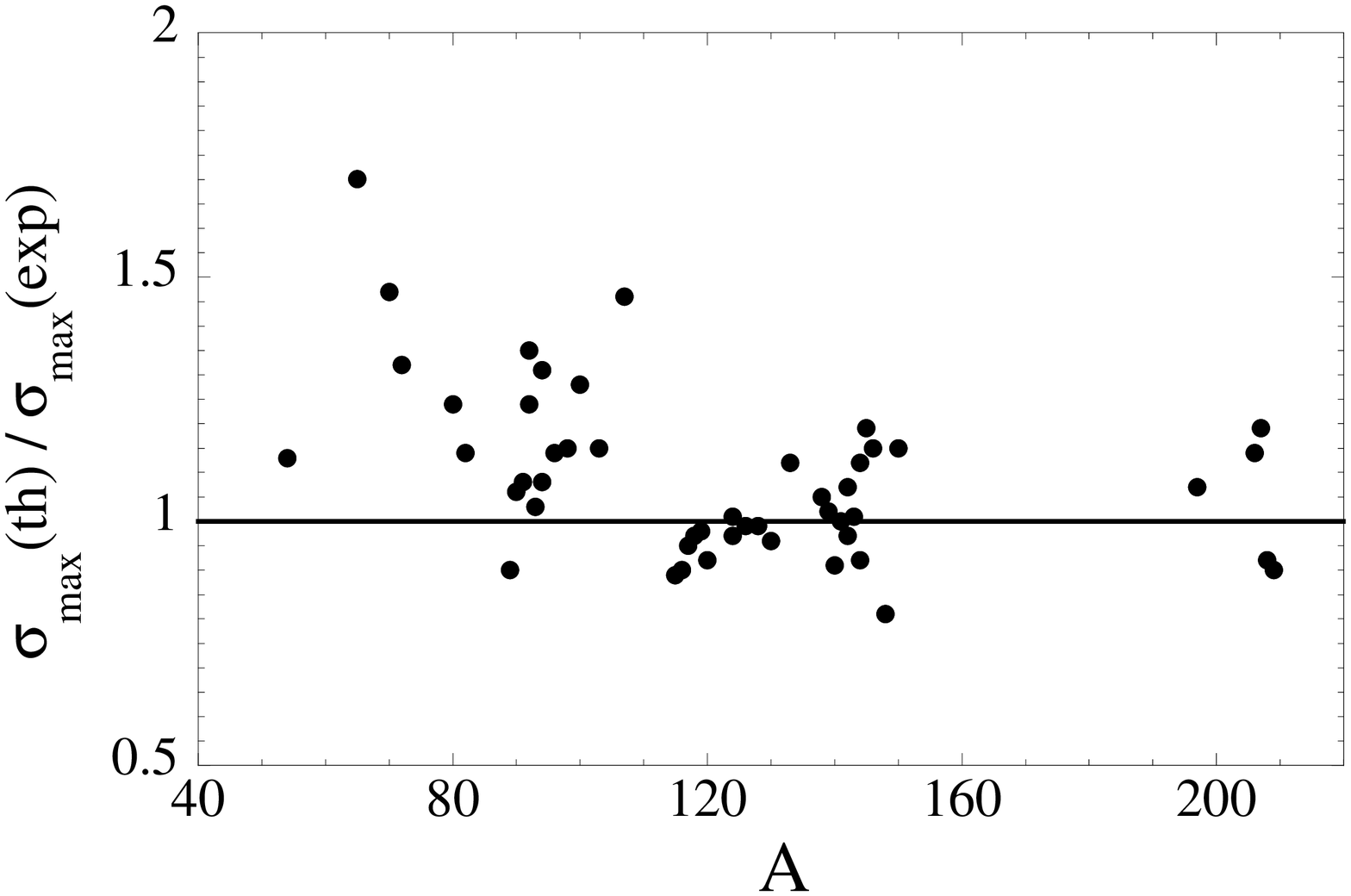,height=9.0cm,width=14cm}}
\caption{Ratio of the peak cross section $\sigma_{max}$(th) estimated within the HFB+QRPA
model with the BSk7 Skyrme force to the experimental value $\sigma_{max}$(exp) for the 48
spherical nuclei as a function of the mass number $A$.}
\label{fig_sigmax}
\end{figure}

Finally, regarding the amplitude of the E1-strength function, the QRPA
equations are solved so as to exhaust the Thomas-Reiche-Kuhn sum rule.  This
adopted EWSR corresponds to
\begin{equation}
\int_0^{\omega_{max}} \sigma(E)~dE ~=~60~NZ/A~[{\rm MeV~b}]
\end{equation}
and is found to reproduce  well the peak cross section measured experimentally, as
illustrated in Fig.~\ref{fig_sigmax}. The resulting deviation can be
characterized by an rms deviation factor $f_{rms}=1.18$ defined as
\begin{equation}
f_{rms}={\mathrm exp} \left[\frac{1}{N_e} \sum_{i=1}^{N_e} \ln^2
\frac{\sigma^i_{max}(th)}{
\sigma^i_{max}(exp)} \right]^{1/2}
\label{eq_rms}
\end{equation}
\noindent where $\sigma_{max}(th)$($\sigma_{max}(exp)$) is the theoretical
(experimental) peak cross section and $N_e=48$ the number of nuclei in the
experimental sample.

All these results show that among the six Skyrme forces studied
here, both the BSk6 and BSk7 forces not only reproduce extremely well the experimental
masses (with a rms deviation as low as 0.676~MeV on the 2135 known masses), but also is
well adapted to describe the E1 collective excitations. For this reason, all further
calculations are performed with BSk7 as our standard force. It should be stressed that
for more than thirty years, phenomenological effective interactions were developed using
exclusively ground state properties of nuclei, such as binding energies, radii or
spectroscopic quantities. This was initiated through the Skyrme Hartree-Fock model by
\cite{va72}. Nuclear forces are traditionally determined by fitting such ground state
properties for less than ten or so nuclei. Recently, progress has been achieved in
determining the Skyrme force by fitting essentially all the mass data
\cite{ms02,sg02,ms03,ms03b}. The only excited feature taken into account so far was the
giant monopolar resonance energy \cite{ng81} in order to predict the infinite
matter compressibility modulus. The present HFB+QRPA model (with the SRPA corrections)
allows to consider nuclear excitations such as GDR in the development of
phenomenological effective interactions.

\subsection{Generalization to deformed nuclei}

\begin{figure}
\centerline{\epsfig{figure=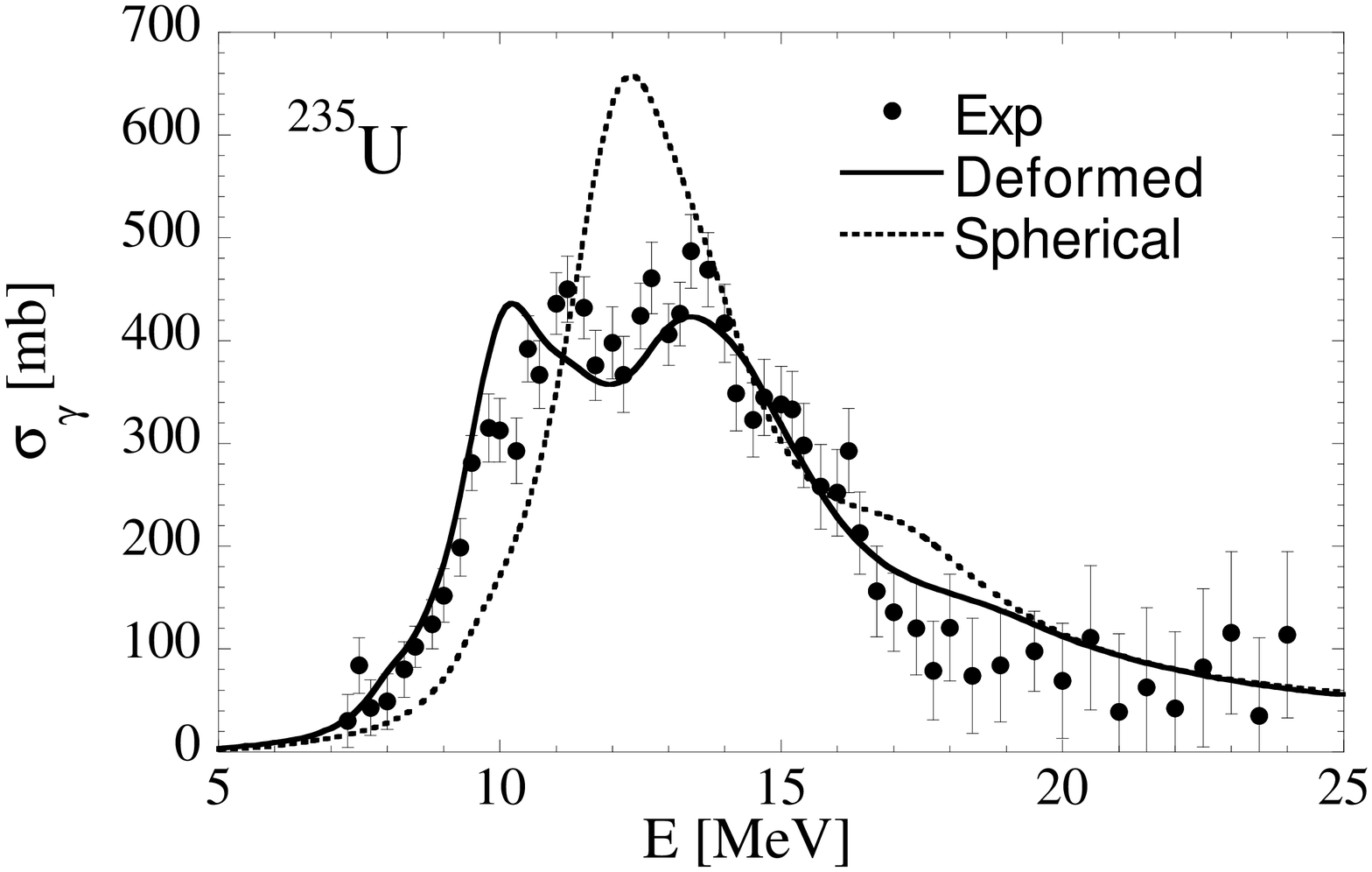,height=9.0cm,width=14cm}}
\caption{Photoabsorption cross section for $^{235}$U. The dots correspond to
experimental data \cite{iaea00}. The dotted line is the HFB+QRPA calculation obtained
with the BSk7 force in the spherical approximation (applying the SRPA) and the full line
when applying in addition our phenomenological procedure to describe deformation
effects.}
\label{fig_u235}
\end{figure}

In the case of deformed spheroidal nuclei, the GDR splits into two major
resonances as a result of the different resonance conditions characterizing the
oscillations of protons against neutrons along the axis of rotational symmetry and an
arbitrary axis perpendicular to it. In the phenomenological approach, the Lorentzian-type
formula is generalized to a sum of two Lorentzian-type functions of energies $E^l_{GDR}$
and width $\Gamma_{GDR}^l$ \cite{th83}, such that
\begin{eqnarray}
\label{eq_def}
& & E^1_{GDR}+2~E^2_{GDR}=3E_{GDR} \\ \nonumber
& & E^2_{GDR}/E^1_{GDR}=0.911 \eta + 0.089
\end{eqnarray}
where $\eta$ is the ratio of the diameter along the axis of symmetry to the diameter
along an axis perpendicular to it. In turn, the width
$\Gamma_{GDR}^l$ of each peak is given by the same deformation dependence as the 
respective energy $E^l_{GDR}$ \cite{th83}. A similar splitting of the resonance strength
for deformed nuclei is applied within the SRPA procedure given by Eq.~(\ref{loren}), 
the Lorentzian function at a given energy $E'$
 being split with an equal strength into two Lorentzian functions centered according
to Eq.~(\ref{eq_def}) and characterized by a width $\Gamma(E)$ (see Eq.~\ref{width})
obtained from the same relations~(Eq.~\ref{eq_def}). As already found in
\cite{sg02}, distributing the strength equally between the two resonance peaks gives
optimal location and relative strength of both GDR centroid energies as observed
experimentally.  We illustrate in Fig.~\ref{fig_u235} how the photoabsorption cross
section in $^{235}$U peaked around 12~MeV in the spherical approximation is split into
the two observed peaks. The same deformation effects are applied to all nuclei predicted
to be deformed by the HFB calculation based on the BSk7 force.

\subsection{Low-energy E1-strength data}

For practical astrophysics applications, it is of first importance to describe the tail
of the GDR  at low energies, i.e around the neutron separation energy, as reliably as
possible \cite{go98}. Experimental E1 strengths at low energies are available through
average resonance capture (ARC) data \cite{ko90} or recent measurements of $\gamma$-ray
spectra in light-ion reactions \cite{voi01,siem02}. However, such data are related to the
so-called ``downwards'' E1-strength function which determines the average width of the
$\gamma$-decay, while the photoexcitation data considered so far depend on the
``upwards'' E1-strength function associated with $\gamma$-absorption. When dealing
with $\gamma$-decay data, a temperature-dependent correction factor is traditionally
introduced in the expression of the GDR width to take the collision of
quasiparticles into account \cite{ka83,ko90,bo01}. In order to guarantee the
compatibility with photoabsorption data, we introduce in the SRPA procedure such a
collision term by adding to the width $\Gamma(E)$ (see Eq.~\ref{width}) a
temperature-dependent correction term as
\begin{equation}
\Gamma'(E)=\Gamma(E) [ 1 + \alpha \frac{4\pi T^2}{E E_{GDR}} ]
\label{tdep}
\end{equation}
where $T$ refers to the temperature of the absorbing state, $E_{GDR}$ is the peak energy
of the GDR and $\alpha$ a normalization constant. In all calculations performed in the
present work, the temperature is derived from the microscopic statistical model of
nuclear level densities \cite{dem01}. As shown below, adopting $\alpha=3$ gives
excellent agreement with most of the available data. 

\begin{figure}
\centerline{\epsfig{figure=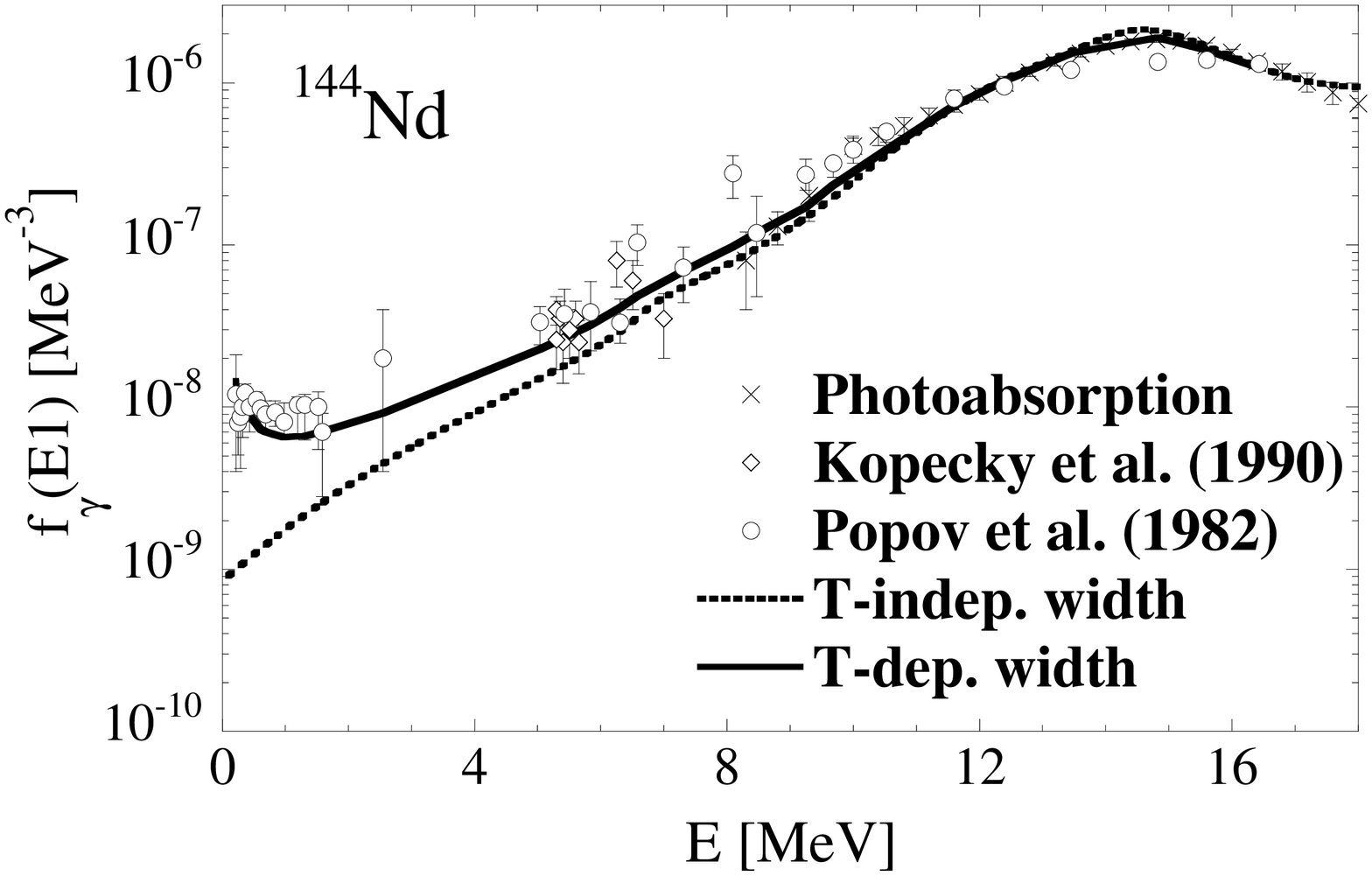,height=9.0cm,width=14cm}}
\caption{\label{fig_nd144}{Comparison of the photoabsorption data \cite{di89} and
measured primary photon strength functions for (n,$\gamma$)
reaction (crosses) \cite{ko90} and  (n,$\gamma\alpha$) reaction (circles) \cite{pop82}
with the QRPA predictions obtained with a $T$-independent (Eq.~\ref{width}) and
$T$-dependent  (Eq.~\ref{tdep}) width. The QRPA predictions are obtained with the BSk7
Skyrme force and a temperature $T=0.55$~MeV.}}
\end{figure}

\begin{figure}
\centerline{\epsfig{figure=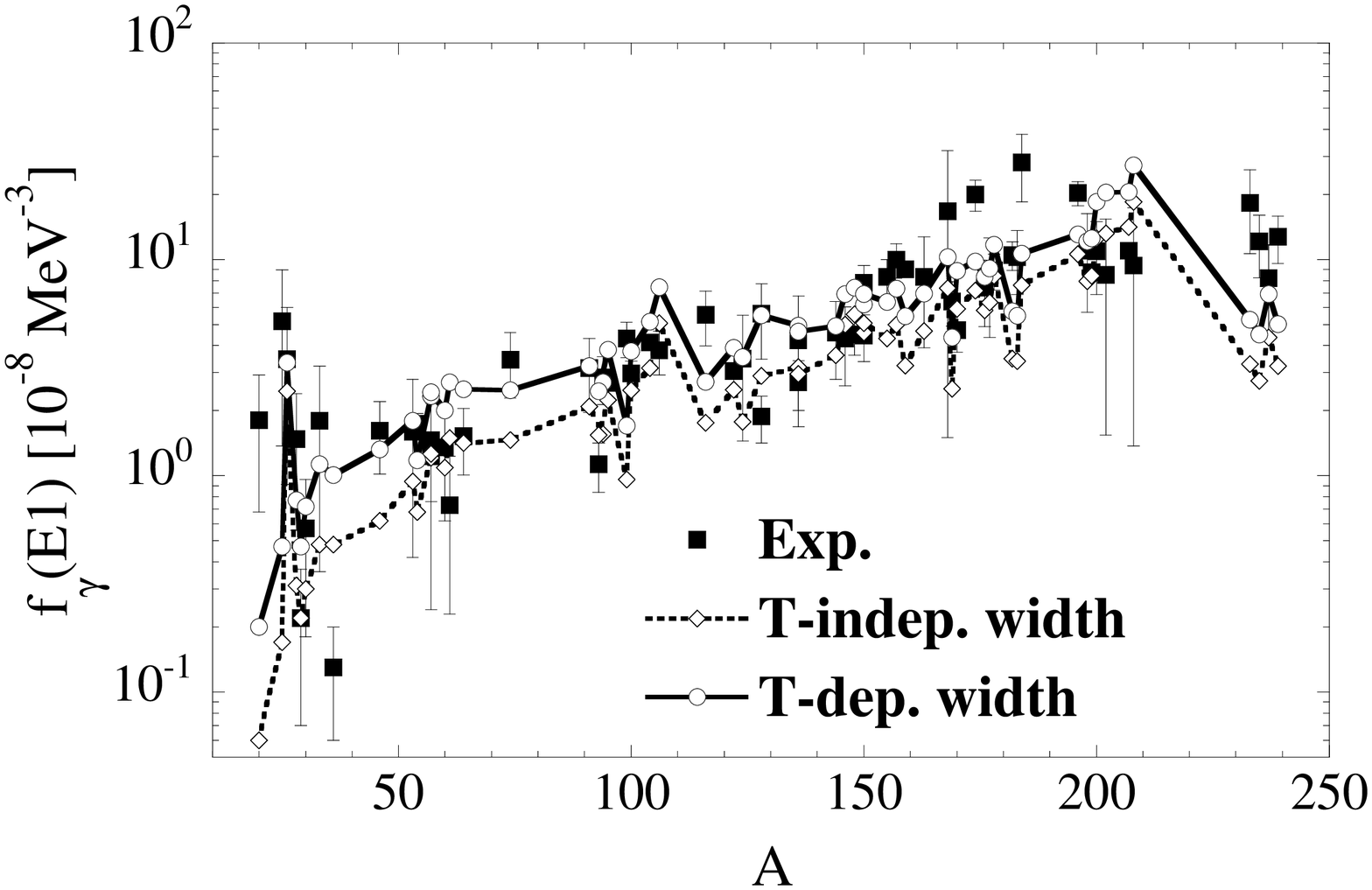,height=9.0cm,width=14cm}}
\caption{\label{fig_fe1}{Comparison of the QRPA $T$-dependent and $T$-independent
low-energy
E1-strength functions with the experimental compilation
\cite{ripl2} including resolved-resonance and thermal-captures measurements, as well as
photonuclear data for nuclei from $^{25}{\rm Mg}$ up to $^{239}{\rm U}$ at energies
ranging from 4 to 8 MeV. }}
\end{figure}

Fig.~\ref{fig_nd144} illustrates in the specific case of the spherical
$^{144}$Nd nucleus, that the E1-strength data derived from primary photon
spectra in the (n,$\gamma$) reaction around 6--8 MeV
\cite{ko90} or  (n,$\gamma\alpha$) reaction around $E\simeq 1$~MeV \cite{pop82} are
correctly reproduced at low energies with the $T$-dependent correction given by
Eq.~(\ref{tdep}) with $\alpha=3$. The energy dependence of the collision term introduced
in Eq.~(\ref{tdep}) and already suggested in
\cite{go98} is of particular importance, since it is responsible for the
$E\rightarrow 0$ behavior of the E1-strength function observed experimentally
\cite{ko90}. It is also found to affect the E1 strength around the neutron
binding energy, as seen in Figs.~\ref{fig_nd144}--\ref{fig_fe1}. 

In addition, we compare in Fig.~\ref{fig_fe1} the QRPA predictions with the
compilation of experimental E1-strength functions at low energies ranging from 4 to 8 MeV
\cite{ripl2} for nuclei from $^{25}{\rm Mg}$ up to $^{239}{\rm U}$. The data set
includes resolved-resonance measurements, thermal-captures measurements and photonuclear
data. In a certain number of cases the original experimental values need to be corrected,
typically for non-statistical effects, so that only values
recommended by \cite{ripl2} are considered in Fig.~\ref{fig_fe1}. QRPA predictions are
 globally in good agreement with experimental data at low energies in the
whole nuclear chart. The average and rms deviations, as defined in Eq.~(\ref{eq_rms}), on
the 62 experimental data have been estimated. The $T$-independent predictions
underpredict the E1 strength by an average factor of 1.6, while on average the
$T$-dependent formula (assuming $\alpha=3$ in Eq.~\ref{tdep}) is in perfect agreement
with the data. The respective rms deviation factors are $f_{rms}=2.6$ and 2.1 for the
$T$-independent and $T$-dependent results. These results show that including a
$T$-dependence in the E1 strength to describe the
$\gamma$-decay data globally improves the agreement. A qualitative agreement is also
obtained with the E1-strength function derived at low energy from primary photon spectra
in light-ion reactions \cite{voi01,siem02}, although the fine structure pattern are not
reproduced.
 
\section{Extrapolation to neutron-rich nuclei and application to the radiative neutron
capture}
\begin{figure}
\centerline{\epsfig{figure=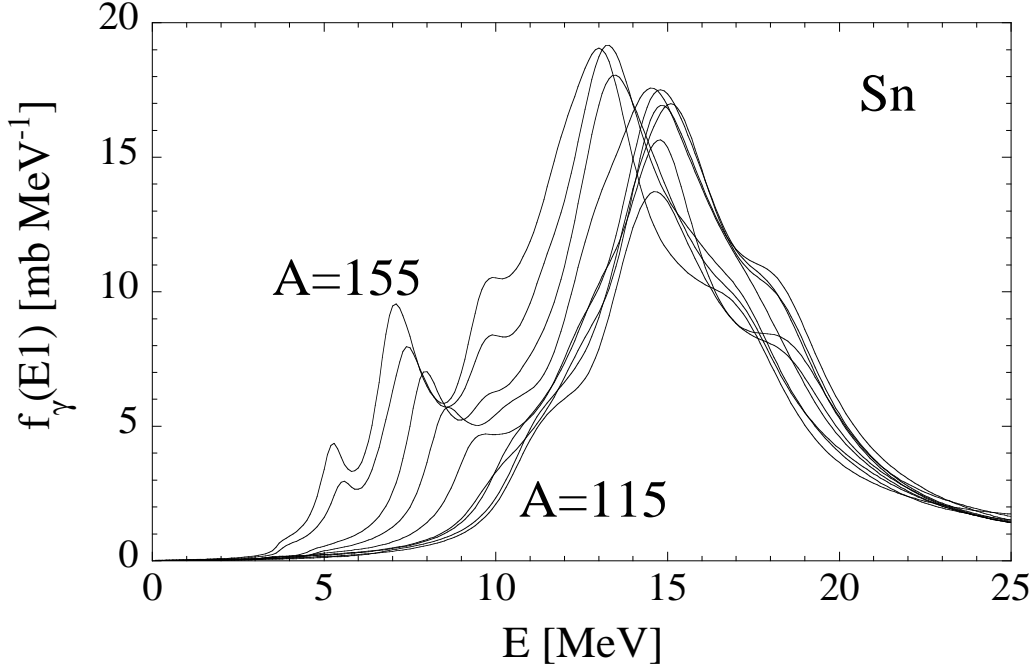,height=9.0cm,width=14cm}}
\caption{\label{fig_sn}{E1-strength function for the Sn
isotopic chain predicted by the HFB+QRPA with the BSk7 force. The SRPA is applied to
all distribution. Only isotopes ranging between A=115 and A=150 by steps of $\Delta A$=5
are displayed.}}
\end{figure}

Large-scale QRPA calculations based on the BSk7 Skyrme force
have been performed for all $8 \le Z\le 110$ nuclei lying between the proton
and the neutron driplines, i.e some 8300 nuclei. The SRPA is applied to all
distributions. In the neutron-deficient region, as well as along the valley of
$\beta$-stability, the resulting E1-strength functions are very similar to
the empirical Lorentzian-like approximation. When dealing with
neutron-rich nuclei, the QRPA predictions start deviating from a simple
Lorentzian shape and results quantitatively similar to \cite{go02} are obtained. In
particular, some extra strength is found to be located at an energy lower than the GDR
energy. The more exotic the nucleus, the stronger this low-energy component. This is
illustrated in Fig.~\ref{fig_sn} for the E1-strength function in the Sn isotopic chain.
All nuclei shown in Fig.~\ref{fig_sn} are predicted to be spherical in the HFB
calculations based on the BSk7 force \cite{ms03b}. For the $A\ge 140$ neutron-rich
isotopes, an important part of the strength is concentrated at low energies ($E \lsimeq
5-7$~MeV). Phenomenological models are unable to
predict such low energy components. In
particular for $^{150}$Sn, all phenomenological systematics (as used for
cross section calculation) predict a $\gamma$-ray strength peaked around
15~MeV with a full width at half maximum of about 4.5~MeV
\cite{ripl2} which is obviously very different from the microscopic estimate
(Fig.~\ref{fig_sn}). 
More generally, the present HFB+QRPA calculation confirms that the
neutron excess affects the spreading of the isovector dipole strength, as well as the
centroid of the strength function. The energy shift is larger than predicted
by the usual $A^{-1/6}$ or $A^{-1/3}$ dependence given by the phenomenological liquid
drop approximations \cite{my77}. The above-described feature of the QRPA E1-strength
function for nuclei with a large neutron excess is qualitatively independent of the
adopted effective interaction. 



%

\begin{figure}
\centerline{\epsfig{figure=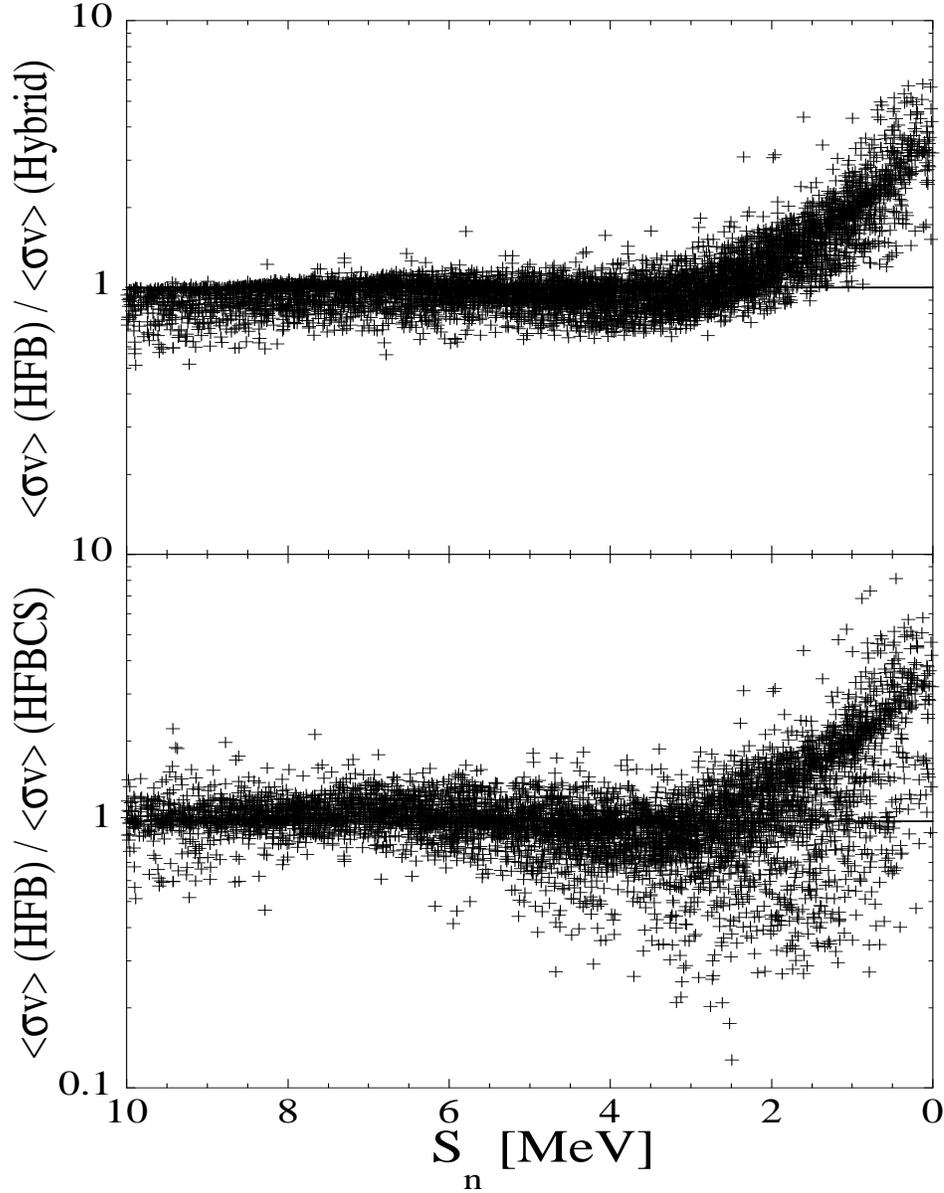,height=16.0cm,width=14cm}}
\caption{\label{fig_rate}{Upper panel: Ratio of the Maxwellian-averaged $(n,\gamma)$ 
rate (at a temperature of $1.5~10^9$~K) obtained
with the HFB+QRPA E1 strength to the one using the Hybrid formula
\cite{go98} as a function of the neutron separation
energy $S_n$ for all nuclei with $8\le Z \le 110$. Lower panel: Same as upper panel 
where the
HFB+QRPA neutron capture rates are compared with the HFBCS+QRPA rates of \cite{go02}.}}
\end{figure}

The radiative neutron capture cross section is estimated within the
statistical model of Hauser-Feshbach making use of the MOST code
\cite{go01a}. It should be noted that this version makes use of the nuclear
ground state properties derived coherently from the same microscopic HFB
method with the BSk7 Skyrme force
\cite{ms03b}. It also benefits from the improved nuclear level density prescription based
on the microscopic statistical model, also used to estimate the nuclear
temperature in Eq.~(\ref{tdep}) \cite{dem01}. The direct capture
contribution as well as the possible overestimate of the statistical
predictions for resonance-deficient nuclei are effects that could have an
important impact on the radiative neutron captures by exotic nuclei
\cite{go98}, but are not included in the present study. The
Maxwellian-averaged radiative neutron capture rate at a temperature
$T=1.5~10^9$~K, typical of the r-process nucleosynthesis, obtained with the
QRPA E1-strength are compared in Fig.~\ref{fig_rate} with those based on the
Hybrid Lorentzian-type formula
\cite{go98}. These rates are sensitive to the neutron capture cross section at incident
energies around 130 keV, and therefore depend on the E1 strength in a narrow range of a
few hundred keV around $S_n$. The temperature-dependent Hybrid formula corresponds to a
generalization of the energy- and temperature-dependent Lorentzian formula including an
improved description  of the E1-strength function at energies below $S_n$ as derived from
\cite{ka83}. The Hybrid E1 strength differs from the QRPA estimate not only
in the location of the centroid energy, but also in the low-energy tail. No extra
low-lying strength is included in the phenomenological Hybrid formula, but its
temperature dependence increases the E1 strength at low energies and is responsible for
its non-zero $E \rightarrow 0$ limit. The newly-derived strength gives an increase of
the rate by a factor up to 6 close to the neutron drip line. R-process nuclei
characterized by
$S_n \lsimeq 3$~MeV are seen to have a neutron capture rate about at least twice faster
than the one predicted with the phenomenological Hybrid formula. This is due to the
shift of the GDR to lower energies compared with the usually adopted
liquid-drop $A^{-1/3}$ rule, as well as to the appearance of some extra strength at low
energies as explained above. Both effects tend to increase the E1 strength at energies
below the GDR, i.e in the energy window of relevance in the neutron capture process. For
less exotic nuclei, the QRPA impact is relatively small, differences being mainly due to
the exact position of the GDR energy and the resulting low-energy tail. When compared to
our previous HFBCS+QRPA predictions \cite{go02}, the
HFB+QRPA model gives larger neutron capture rates close to the neutron drip line, but
lower rates for many of the $4\lsimeq S_n~[{\rm MeV}] \lsimeq 2$ nuclei, as seen in
Fig.~\ref{fig_rate} (lower panel). These differences justify the use of the HFB approach
for exotic neutron-rich nuclei. 

\section{Conclusions}
The E1-strength function is estimated with one of the most accurate and reliable
microscopic model available to date, namely the Hartree-Fock-Bogoliubov (HFB) and QRPA
models. The spreading width of the GDR is taken into account by analogy with the
SRPA method. The analysis of HFB+QRPA model based on various Skyrme forces with different
pairing prescriptions and parameterizations shows that the effective nucleon-nucleon
interaction can be constrained with the GDR data. In particular, it is found that the
Skyrme force characterized with a low effective mass $M^*_s/M\simeq 0.8$ is a necessary
condition to reproduce the location and width of the GDR, at least within the present
HFB+QRPA model to which the SRPA is applied. In contrast, GDR data cannot be used to
discriminate between the surface or volume property of the pairing interaction. In
addition to its reliability, it is shown that the HFB+QRPA model also gives accurate
predictions and that globally it agrees fairly well with experimental data.
The present HFB+QRPA model brings important improvement with respect to our previous
HFBCS+QRPA model and can provide quantitative predictions of the dipole strength.
Large-scale calculations of the E1-strength function are performed and used to estimate
the radiative neutron capture rates of relevance for the r-process nucleosynthesis. A
systematic increase of the reaction rates for exotic neutron-rich nuclei is found.

Further improvements may be useful. A proper treatment of the continuum
states and its impact on the dipole strength is an important issue. It is expected to be
significant for drip-line nuclei. Continuum-QRPA models are available \cite{kh02} and
study along these lines are in progress. The particle-vibration coupling also
affects the low-energy strength and could contribute to an extra increase of the
radiative neutron capture rate by exotic nuclei.

\smallskip
\noindent{\bf Acknowledgments} M.S. and S.G. are FNRS Research Fellow and Associate,
respectively. This work has been performed within the  scientific collaboration
(Tournesol) between the Wallonie--Bruxelles Community and France. \\

\end{document}